\documentclass[12pt]{article}
\usepackage{graphicx,amssymb,epsfig}
\begin{document}
\begin{titlepage}
\begin{center}
{\large Energy Spectrum of a Relativistic Two-dimensional \\
Hydrogen-like Atom in a Constant  \\
Magnetic Field of arbitrary strength}

\vspace{4ex}

V\'{\i}ctor M. Villalba\footnotemark[1],
Ramiro Pino\footnotemark[2]
\vspace{4ex}

{\footnotemark[1]\footnotesize\it Centro de F\'{\i}sica, Instituto Venezolano de Investigaciones Cient\'{\i}ficas, IVIC \\
Apdo 21827, Caracas 1020-A, Venezuela}

\vspace{4ex}

{\footnotemark[2]\footnotesize\it 
Department of Mathematics and Computing Science \\
Technische
Universiteit Eindhoven, \\
P. O. Box 513 Eindhoven 5600 MB, The Netherlands} 

\footnotetext[1]{e-mail: villalba@ivic.ve}
\footnotetext[2]{e-mail: rpino@win.ue.nl}
\end{center}

\begin{abstract}
We compute, via a variational mixed-base method, the energy spectrum of a
two dimensional relativistic atom in the presence of a constant magnetic
field of arbitrary strength. The results are compared to those obtained in
the non-relativistic and spinless case. We find that the relativistic
spectrum does not present $s$ states.
\end{abstract}
\end{titlepage}

\section{Introduction}
Two-dimensional Hydrogen atoms in magnetic fields have been a subject of
active research during the last years. A large body of articles has been
published on this problem in the framework of non-relativistic quantum
mechanics. This problem is of practical interest because of the technological
advances in nanofabrication technology that have made possible to create low
dimensional structures like quantum wells, quantum wires and quantum dots 
\cite{Greene,Johnson,Chakraborty}. 
The two-dimensional Hamiltonian describing the Coulomb interaction $-\frac
Zr$, between a conduction electron and donor impurity center when a constant
magnetic $\vec {{B}}$ field is applied perpendicular to the plane of
motion, can be written in atomic units, $\hbar =M=e=1$ in the CGS
system, as follows 
\begin{equation}
\label{1}H\varphi =\frac 12(-i\nabla +\frac 12\vec {{B}}\times \vec
r)^2\varphi -\frac Zr\varphi =i\partial _t\varphi =E\varphi 
\end{equation}
Since we are dealing with a two-dimensional problem, we choose to work in
polar coordinates $(r,\vartheta ).$ The angular operator operator $%
-i\partial _\vartheta $ commutes with the Hamiltonian (\ref{1}),
consequently we can introduce the following ansatz for the eigenfunction 
\begin{equation}
\label{2}\varphi (\vec r)=\frac{\exp (im\vartheta )}{\sqrt{2\pi }}\frac{u(r)%
}{\sqrt{r}}. 
\end{equation}
Substituting (\ref{2}) into (\ref{1}), we readily obtain that the radial
function $u(r)$ satisfies the second order differential equation

\begin{equation}
\label{ecua}\left[ -\frac 12\frac{d^2}{dr^2}+\frac 12(m^2-\frac 14)\frac
1{r^2}+\frac{\omega _L^2r^2}2-\frac Zr+m\omega _L-E\right] u(r)=0,
\end{equation}
where $\omega _L={B}/2c$ is the Larmor frequency, $E$ is the energy,
and $m$ the eigenvalue of the angular momentum. Eq. (\ref{ecua}) cannot be
solved in closed form in terms of special functions \cite{Bagrov}. There are
analytic expressions for the energy for particular values of $\omega _L$ and 
$m$\cite{Lozanskii,Taut1,Taut2}. The computation of the energy eigenvalues
in (\ref{ecua}) has been carried out using different techniques \cite{MacDonald,Martin,Imbo,Mustafa1,Mustafa2,Villalba2}.
A less studied problem is that of a relativistic 2D hydrogen atom in a
magnetic field. Perhaps relativistic effects are not very important in
semi-conducting devices but nevertheless they cannot be neglected when the
interacting potentials are strong \cite{Elliot,Spector}. Recently,  the importance of 
considering relativistic effects has been pointed out when one computes the energy levels of
semiconductors in high magnetic fields\cite{Avetisyan}. 
The effective mass method is still valid until $H\approx 10^5$ Oe. In the case of two band 
approximation, the dispersion law has the form of a Klein-Gordon energy 
spectrum\cite{Keldysh}.  Also, relativistic invariance
imposes some supplementary restrictions on the allowed quantum energy levels.

In this article we investigate the relativistic corrections to the energy 
spectrum of a two-dimensional hydrogen atom in an homogeneous
transverse magnetic field. 
Using a mixed-basis variational approach \cite{Chen1,Chen2}. In Sec 2, we compute the relativistic energy
spectrum of a 2D relativistic Klein-Gordon hydrogen atom.
In Sec
3, we discuss the application of the $1/N$ expansion to our problem. In Sec 4, we
compare the energy spectrum of the relativistic 2D hydrogen atom with that
obtained in the nonrelativistic limit. Finally, we present the concluding
remarks in Sec. 5. 

\section{Relativistic Hydrogen Atom}

Since we are interested in discussing the relativistic corrections to the
energy levels of the 2D Hydrogen atom, we proceed to solve the 2D Klein-Gordon
equation. The results obtained after solving the Klein-Gordon equation apply 
to an electron without spin.  The advantage of this approach \cite{Spector} can be easily understood if we recall that the Sch\"odinger equation does not take into account the
spin of the electron (\ref{1}) and then we can directly compare the
relativistic and nonrelativistic energy spectra. 

The covariant generalization of the Klein-Gordon equation in the presence of
electromagnetic interactions takes the form \cite{Villalba2,Davydov} 
\begin{equation}
\label{3}\left( g^{\alpha \beta }(\nabla _\alpha -\frac icA_\alpha )(\nabla
_\beta -\frac icA_\beta )-c^2\right) \Psi =0, 
\end{equation}
where $g^{\alpha \beta }$ is the contravariant metric tensor, and $\nabla
_\alpha $ is the covariant derivative. The metric tensor $g_{\alpha \beta }$
written in polar coordinates $(t,r,\vartheta )$ takes the form: 
\begin{equation}
\label{4}g_{\alpha \beta }=diag(-1,1,r^2), 
\end{equation}
and the vector potential $A^\alpha $ associated with a 2D Coulomb potential and a
constant magnetic field interaction is 

\begin{equation}
\label{5}A^\alpha =(-\frac Zr,0,-\frac{{B}r^2}2). 
\end{equation}
From the above expression (\ref{5}) for the vector potential $A^\alpha $ it
is straightforward to verify that the electric and magnetic fields satisfy
the invariant relations 
\begin{equation}
\label{seis}F_{\alpha \beta }F^{\alpha \beta }=2({B}^2-{E}^2)=2(%
{B}^2-\frac{Z^2}{r^4}), 
\end{equation}
\begin{equation}
\label{siete}^{*}F_{\alpha \beta }F^{\alpha \beta }=0\ \rightarrow \vec {%
{E}}\cdot \vec {{B}}=0, 
\end{equation}
where $F^{\alpha \beta }$ is the (2+1) electromagnetic field strength tensor.

Expressions (\ref{seis}) and (\ref{siete}) tell us that in fact, $A^{\alpha }$
is associated with a 2D Coulomb atom in a constant magnetic field
perpendicular to the plane of the particle motion. The corresponding $%
\vec{{E}}$ and $\vec{{B}}$ can be written in polar coordinates as
follows: 
\begin{equation}
\label{E}\vec{{E}}=-\frac{Z}{r^{2}}\hat{e}_{r},\quad \vec{{B}}=%
{B}\hat{e}_{z}. 
\end{equation}
Since the vector potential components do not depend on time or the angular
variable $\vartheta ,$ we have that the wave function $\Psi $, solution of
the Klein-Gordon equation (\ref{3}), can be written as

\begin{equation}
\label{6}\Psi (r,\vartheta ,t)=\frac{u(r)}{\sqrt{r}}\exp (im\vartheta -Et), 
\end{equation}
where the function $u(r)$ satisfies the second order differential equation

\begin{equation}
\label{7}\frac{d^2u(r)}{dr^2}+\left( \frac{\frac 14-m^2+\frac{Z^2}{c^2}}{r^2}%
-\frac{m{B}}c-c^2+\frac{E^2}{c^2}-\frac 14\frac{r^2{B}^2}{c^2}+%
\frac{2EZ}{c^2r}\right) u(r)=0. 
\end{equation}
Eq. (\ref{7}) has the same form as Eq. (\ref{ecua}), therefore no exact
solutions of Eq. (\ref{7}) can be obtained in terms of special functions.

In the present article we analyze the problem of computing the energy levels
of the 2D relativistic Coulomb atom using a mixed-basis variational
approach. In order to apply the variational method to our problem \cite
{Davydov}, we look for a trial wave function. Since Eq. (\ref{7}) reduces to
the relativistic Hydrogen atom equation when $\omega _L=0,$ we can consider
as a basis, for $\omega _L<<1$, the Hydrogen wave functions $u_H.$ The
solution of eq. (\ref{ecua}) when $\omega _L=0$ is

\begin{equation}
u_H(r)=D_{m,n}e^{-r\sqrt{c^2-\frac{E^2}{c^2}}}r^{(\sqrt{m^2-\frac{Z^2}{c^2}}%
+1/2)}L(n_\rho ,2\sqrt{m^2-\frac{Z^2}{c^2}},2r\sqrt{c^2-\frac{E^2}{c^2}})
\end{equation}
where $D_{m,n}$ is a normalization constant,  $L(a,b,x)$ are the Laguerre
polynomials \cite{Lebedev}, and $E$ from \cite{Nieto} is 
\begin{equation}
\label{8}E=c^2\left[ 1+\frac{Z^2}{c^2(n_\rho -\frac 12+\sqrt{m^2-\frac{Z^2}{%
c^2}})^2}\right] ^{-1/2}.
\end{equation}
It is worth mentioning that the relation (\ref{8}) makes sense only when 
\begin{equation}
\label{ineq}m^2-\frac{Z^2}{c^2}>0,
\end{equation}
a condition that forbids the existence of the $s$ energy levels ($m=0)$,
this is in fact a peculiarity of the relativistic Klein-Gordon solution,
which is not present in the standard Schr\"odinger framework.

Conversely, for large values of $\omega _L,$ a good trial basis is that of
the spherical oscillator. In this case, the solution of eq. (\ref{7}) has the
form 
\begin{equation}
u_{Osc}(r)=C_{m,n}e^{-\omega _L\rho ^2/2}\rho ^{(\left| m\right|
+1)}L(n_\rho ,\left| m\right| ,\omega _L\rho ^2)
\end{equation}
and, in the high-field limit, the energy spectrum of a relativistic spinless
particle in a constant magnetic field satisfies the relation

\begin{equation}
\label{9}\frac{E^2}{c^2}-c^2=2\omega _L\left( 2n+m+\left| m\right| +1\right). 
\end{equation}

Among the advantages of considering a relativistic spinless electron is that
we can easily compute the energy levels with the help of the mixed base variational approach. Nevertheless, we can easily see when the Klein Gordon equation gives a reasonably good value for the energy spectrum as compared to that obtained via the Dirac equation. 

The 2+1 Dirac equation \cite{shishkin,Khalilov} in the presence of an external
electromagnetic field $A_{\mu}$ reads 
\begin{equation}
\label{Dirac}
\left( \gamma ^{\mu }(\frac{\partial }{\partial x^{\mu }}+\frac{A_{\mu }}{c}%
)+c\right) \Psi =0.
\end{equation}
Since we are working in a two dimensional space, we can work in the following representation of the gamma matrices 
\begin{equation}
\gamma ^{0}=i\sigma ^{3},\quad \gamma ^{1}=\sigma ^{1},\quad \gamma
^{2}=\sigma ^{2}.
\end{equation}
Then, the Dirac spinor has only two components. Since the Dirac
equation (\ref{Dirac}) expressed in the diagonal tetrad gauge commutes with the operators
$i\frac{\partial}{\partial t}$ and $-i\frac{\partial}{\partial \vartheta}$, the spinor $\Psi$ can be written as 
\begin{equation}
\Psi (t,\mathbf{r})=\frac{1}{\sqrt{2\pi }}\exp (-iEt+il\vartheta )\psi,
(r)
\end{equation}

\begin{equation}
\psi =\left(
\begin{array}{c}
\psi _{1}(r) \\
\psi _{2}(r)
\end{array}
\right)
\end{equation}
where $\psi_{1}$ and $\psi_{2}$ satisfy the system of equations
\begin{equation}
\label{eq1}
(\frac{E}{c}+c+\frac{Z}{\rho })\psi _{1}+(\frac{\partial }{\partial \rho }+%
\frac{l}{\rho }+\frac{B\rho }{2c})\psi _{2}=0,
\end{equation}
\begin{equation}
\label{eq2}
(-\frac{E}{c}+c-\frac{Z}{\rho })\psi _{2}+(\frac{\partial }{\partial \rho }-%
\frac{l}{\rho }+\frac{B\rho }{2c})\psi _{1}=0.
\end{equation}
Substituting (\ref{eq1}) into (\ref{eq2}), we obtain the following second order
differential equation
\begin{equation}
\label{seg}
\frac{d^{2}\psi _{2}}{d\rho ^{2}}+\left( (\frac{E}{c}+\frac{Z}{\rho }%
)^{2}-c^{2}-\frac{l(l+1)}{\rho ^{2}}-\frac{1}{4}\frac{B^{2}\rho ^{2}}{c^{2}}%
-(l+\frac{1}{2})\frac{B}{c}+D\right) \psi _{2}=0
\end{equation}
where $D$ is given by 
\begin{equation}
\label{D}
D=\frac{Z}{\rho ^{2}}(\frac{E}{c}+c+\frac{Z}{\rho })^{-1}(\frac{\partial }{%
\partial \rho }+\frac{l}{\rho }+\frac{B\rho }{2c})\psi _{2}
\end{equation}
It is worth mentioning that angular parameter $l$ takes half integer values
\cite{shishkin} and
therefore it can be related to $m$ as follows
\begin{equation}
\label{ele} 
l=m-\frac{1}{2}.
\end{equation}

Eq. (\ref{seg}) reduces to the radial
Klein-Gordon equation (\ref{7}) when $D$ vanishes. Looking at (\ref{D}) and
(\ref{eq1})
we see that $D$ is very small for large values of $\rho$ \cite{Khalilov}. The
mean square radius of the Dirac and Klein-Gordon electron is $\rho^2\approx
2(n+1/2)c/B$ where $n$ labels the energy levels. Taking into account that large
values of the radial variable imply that $\rho>1/c$, we find that  $D$ is
negligible for magnetic
fields satisfying the inequality $B<2c^3$, which is the critical value for strong magnetic
fields \cite{Khalilov2}.

Using Eq. (\ref{seg}), and keeping only  leading terms of $D$, the
motion of a relativistic electron for small values of $B$ and $\rho$ is
described by the equation
\begin{equation}
\label{weak}
\frac{d^{2}\varphi _{2}}{d\rho ^{2}}+\left( (\frac{E}{c}+\frac{Z}{\rho }%
)^{2}-c^{2}+\frac{\frac{1}{4}-l^{2}}{\rho ^{2}}-\frac{1}{4}\frac{B^{2}\rho
^{2}}{c^{2}}-(l-1)\frac{B}{c}\right) \varphi _{2}=0.
\end{equation}
Notice that Eq (\ref{weak}) gives a description of a relativistic electron in
a weak magnetic field $B$ for small values of $\rho$. From Eq. (\ref{seg}) and
(\ref{ele}, we
see that, in oposition to the Klein-Gordon case, the 2+1 relativistic Dirac electron has $s$ states. 

If we attempt to apply the variational method using the hydrogen atom basis,
we will obtain good agreement with accurate results for small values
of $\omega_L,$ but this approach fails for large $\omega_L$,  even if we
consider a many term basis. An analogous situation occurs when we use the
oscillator basis, in which case we obtain a good agreement for large $\omega
_L,$ but the convergence is very slow for small values of $\omega_L $. \cite
{Villalba} In order to solve this problem, we propose a 
trial function \cite{Chen1,Chen2}, for any quantum level $n$, a
linear combination of the form

\begin{equation}
\label{mixed}u=\sum_i^N(c_{iH}u_{iH}+c_{iO}u_{iOsc})
\end{equation}
where $N\geq i\geq n$; $u_{iH}$ and $u_{iOsc}$ are the corresponding
hydrogen and oscillator wave functions associated with the quantum level $i$
 $c_{iO}$ and $c_{iH}$ are constants to be calculated. It is worth
mentioning that our basis is not orthogonal under the inner product $
\left\langle u_i\mid u_j\right\rangle =\int_0^\infty u_iu_jd\rho .$
Substituting (\ref{mixed}) into (\ref{7}), and performing variation on the
basis coefficients $c_j$, we readily obtain the following matrix equation: 
\begin{eqnarray}
\label{matriz} \left(\left\langle u_i(r),\frac{d^2u_j(r)}{dr^2}\right\rangle +(\frac 14-m^2+\frac{Z^2}{c^2})A_{ij}\right) c_j  \\ 
\nonumber
+\left(-  
(\frac{m{B}}c+ 
c^2-\frac{E^2
}{c^2})\delta _{ij}-\frac 14\frac{{B}^2}{c^2}D_{ij}+\frac{2EZ}{c^2}
C_{ij}\right) c_j=0
\end{eqnarray}
with 
\begin{equation}
A_{ij}=\left\langle u_i(r)\left| \frac 1{r^2}\right| u_j(r)\right\rangle ,\\
C_{ij}=\left\langle u_i(r)\left| \frac 1r\right| u_j(r)\right\rangle ,\\
D_{ij}=\left\langle u_i(r)\left| r^2\right| u_j(r)\right\rangle  
\end{equation}
where the indices i and j running from 1 to N correspond to the Hydrogen and
oscillator bases respectively.
The algebraic equation (\ref{matriz}) $Q_{ij}c_j=0$ gives nontrivial values
of $c_j$ provided that the matrix $Q_{ij}$ be singular. The energy
eigenvalue for a given quantum level $n$ is the lowest value of $E,$
solution of the equation $\det (Q_{ij})=0.$
The mixed-basis variational method gives reasonably
good values for the energy eigenvalues even for the simple selection of a
two term basis as in Eq. (\ref{mixed}). In this particular case, we have that the
trial function, for any quantum level, is a linear combination of the form

\begin{equation}
\label{2terms}u_i=c_{iH}u_{iH}+c_{iO}u_{iO} 
\end{equation}
Better results should be expected for a basis with more terms. For a three
term basis, we have two possible trial functions 
\begin{equation}
\label{mix21}u_i=c_1u_{1iH}+c_2u_{2iH}+c_3u_{1iO,}\quad (mix21) 
\end{equation}
and 
\begin{equation}
\label{mix12}u_i=c_1u_{1iH}+c_2u_{1iO}+c_3u_{2iO},\quad (mix12) 
\end{equation}
where the three terms in (\ref{mix21}) and (\ref{mix12}) have the same
angular dependences of the eigenfunction to be approximated. 
In this scheme u$_{2i}$ corresponds to a wavefunction
associated with a higher quantum number to that we are going to approximate.

For comparison, the numerical computations of the relativistic energy
spectra are carried out with the help of the Schwartz method \cite{Schwartz}%
, which is a generalization of the mesh point technique for numerical
approximation of functions. This method gives highly accurate results given
a thoughtful choice of the reference function, and its efficiency has been
shown computing the energy spectrum of the 2D Hydrogen atom \cite{Villalba2}.

\section{1/N Approach}

The shifted 1/N expansion is a perturbative technique that has permitted us to
solve the N-dimensional stationary Schr\"odinger equation with a wide class
of radial potentials. The shifted 1/N method has also been developed to
compute energy eigenvalues of relativistic spin 0 and spin $\frac 12$
particles in the presence of spherically symmetric vector and scalar
potentials. Here we proceed to compute the energy eigenvalues of our problem
using the 1/N expansion for $N=2$. Since Eq. (\ref{7}) contains a magnetic
field contribution, some minor modifications should be made to the recipe
of Ref \cite{Sever} where the authors develop the shifted 1/N
technique to deal with the $3D$ Klein-Gordon equation in a spherically
symmetric potential.

Following the scheme developed by Imbo and Pagnamenta \cite{Imbo}, we have
that the radial Klein-Gordon equation in the presence of a constant magnetic
field (\ref{E}) takes the form 
\begin{equation}
\label{1N}\left( -\frac{d^2}{dr^2}+\frac{(\bar k+a-1)(\bar k+a-3)}{4r^2}%
+mB+M^2+\frac 14r^2B^2-[E+\frac Zr]^2\right) U_{n_r}(r)=0
\end{equation}
where $\bar k=N+2l-a,$ with $a$ as the shifting parameter, and $U_{n_r}(r)$
is the reduced radial wave function. Eq. (\ref{1N}) is written in units $%
\hbar =c=1$. In these units we have that $\alpha =1/137$. Here we borrow the
results reported in Ref \cite{Sever} Introducing the scaled mass 
\begin{equation}
M_a=(M^2+mB)^{1/2}
\end{equation}
and proceeding to minimize $E_0$%
\begin{equation}
E_0=V(r_0)+(M_a^2+Q/(2r_0)^2)^{1/2},
\end{equation}
we obtain that $r_0$ satisfies the equation 
\begin{equation}
r_0^3V^{\prime }(r_0)(1+Q/(4M_a^2r_0^2)^{1/2}=Q/4M_a.
\end{equation}
The shifting parameter is 
\begin{equation}
\label{seize}a=2-(1+2n)w
\end{equation}
where $w$ is given by 
\begin{equation}
\label{dixsept}w=\left( 3+r_0+V^{\prime \prime }(r_0)/V^{\prime
}(r_0)-4r_0^4V^{\prime }(r_0)^2/Q\right) ^{1/2},
\end{equation}
and $Q$ can be written as 
\begin{equation}
\label{dixhuit}Q=\bar k^2,\ Q=[r_o^2V^{\prime }(r_0)]^2(2+2g),
\end{equation}
with 
\begin{equation}
\label{dixneuf}g=(1+(2M_a/(r_oV^{\prime }(r_o)))^{1/2}.
\end{equation}
Equations (\ref{seize}) and (\ref{dixhuit}) along with equations (\ref
{dixsept}) and (\ref{dixneuf}) give 

\begin{equation}
\label{final}2l+(1+2n)w-r_0^2V^{\prime }(r_0)(2+2g)^{1/2}=0.
\end{equation}
Equation (\ref{final}) allows computing the value of $r_0.$ Using this
value we compute the coefficients

\begin{equation}
E_1=E_0V^{\prime \prime \prime }(r_0)-V(r_0)V^{\prime \prime \prime
}(r_0)-3V^{\prime }(r_0)V^{\prime \prime }(r_0), 
\end{equation}
\begin{equation}
E_2=E_0V^{\prime \prime \prime \prime }(r_0)-V(r_0)V^{\prime \prime \prime
\prime }(r_0)-4V^{\prime }(r_0)V^{\prime \prime \prime }(r_0)-3V^{\prime
\prime }(r_0)^2, 
\end{equation}
\begin{equation}
\varepsilon 1=(2-a)/w,\ \varepsilon 2=-3(2-a)/(2w), 
\end{equation}

\begin{equation}
\varepsilon3=-1+r_0^5E_1/(3Qw^{3/2}),\ \varepsilon 4=5/4+r_0^6E_2/(12Qw^2) 
\end{equation}
and finally we have for the energy 

\begin{eqnarray}
E=E_0+1/2/E_0r_0^2((1-a)(3-a)/4+(1+2n)\varepsilon 2+3(1+2n+2n^2)\varepsilon4 \\ \nonumber 
-1/w(\varepsilon1^2+6(1+2n)\varepsilon1\varepsilon3+(11+30n+30n^2)\varepsilon3^2)+B^2r_0^4/4)
\end{eqnarray}
where the energy level is expressed in units where $c=1,$ $\hbar =1.$ The
energy eigenvalues can be expressed in atomic units with the help of the
relation: 
\begin{equation}
E_n=2(E-1)/\alpha ^2.
\end{equation}
Now we proceed to compare the results obtained via the mixed variational
approach as well as the $1/N$ expansion with those computed numerically. Energy
levels are computed in Rydberg units as a function of 
$\gamma^{\prime}=2\omega_L/(2\omega_L+1)$ and displayed in Figures 1 to 4.

\begin{figure}
\epsfig{file=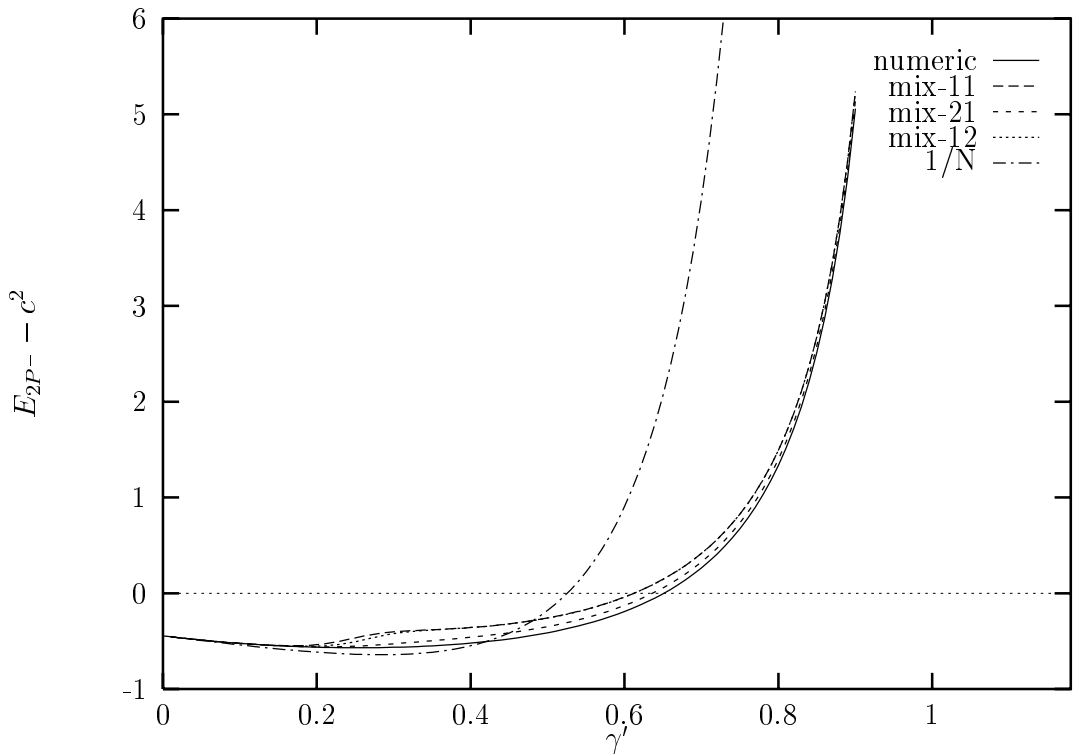,height=10cm}
\caption{Energy of the $2P^-$ state as a function of $\protect\gamma^{\prime
} $. The solid line is obtained by numerical methods; 
the long-dashed line corresponds to the mix11 basis; 
the short-dashed line corresponds to the mix21 basis, ($2P^-$, $3P^-$ Hydrogen bases and $2P^-$ 
oscillator wavefunction). The dotted line is obtained by using the mix12 basis 
($2P^-$, $3D^-$ oscillator bases and the $2P^-$ Hydrogen wavefunction).The dash-dotted  
line is
obtained with the help of the shifted $1/N$ method}.   
\end{figure}

\begin{figure}
\epsfig{file=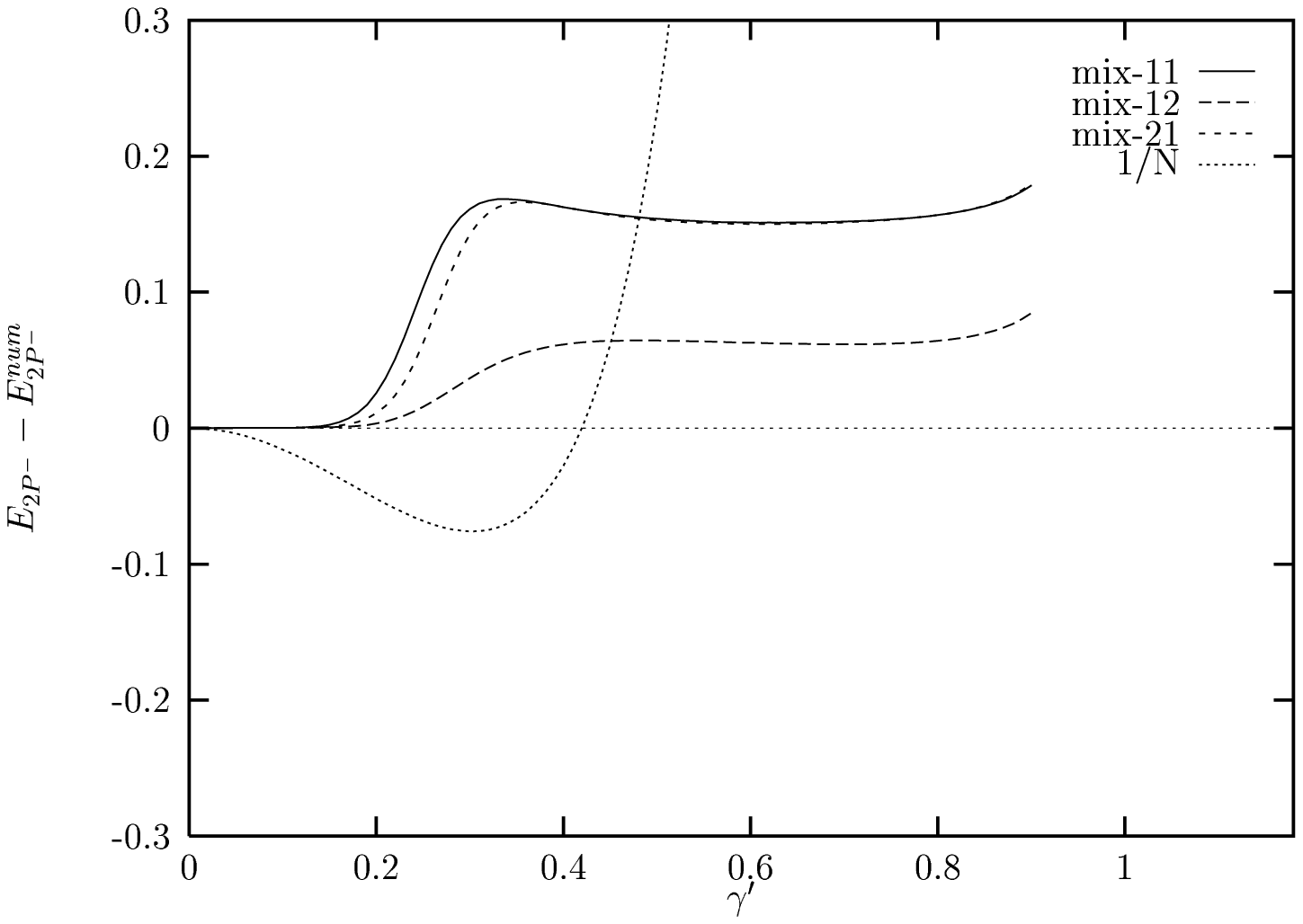,height=10cm}  
\caption{The figure shows the difference between the numeric result for the 
$2P^-$ energy spectrum and the energy values computed with the help of 
the mix11 variational basis (solid line), mix21 variational (light dashed line), 
mix12 variational (heavy dashed line), and the shifted $1/N$ method (dotted line)}
\end{figure}

\begin{figure}
\epsfig{file=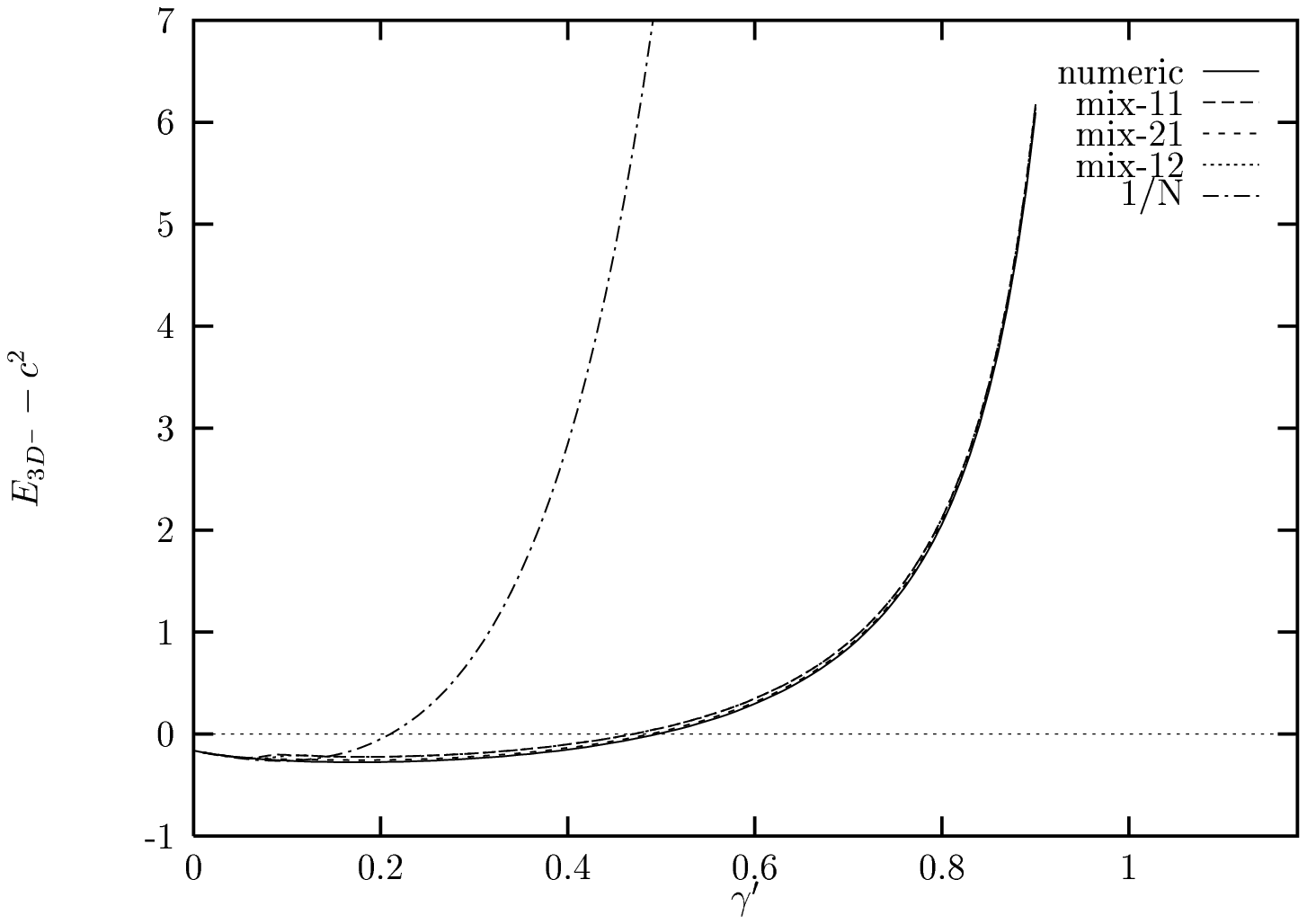,height=10cm}
\caption{Energy of the $3D^-$ state as a function of $\protect\gamma^{\prime
} $. The solid line is obtained by numerical methods; 
the long-dashed line corresponds to the mix11 basis; 
the short-dashed line corresponds to the mix21 basis, ($3D^-$, $4D^-$ Hydrogen bases and $3D^-$ 
oscillator wavefunction). The dotted line is obtained by using the mix12 basis \
($3D^-$, $4D^-$  oscillator bases and the $3D^-$ Hydrogen wavefunction). 
The dash dotted line is
obtained with the help of the shifted $1/N$ method}.
\end{figure}

\begin{figure}
\epsfig{file=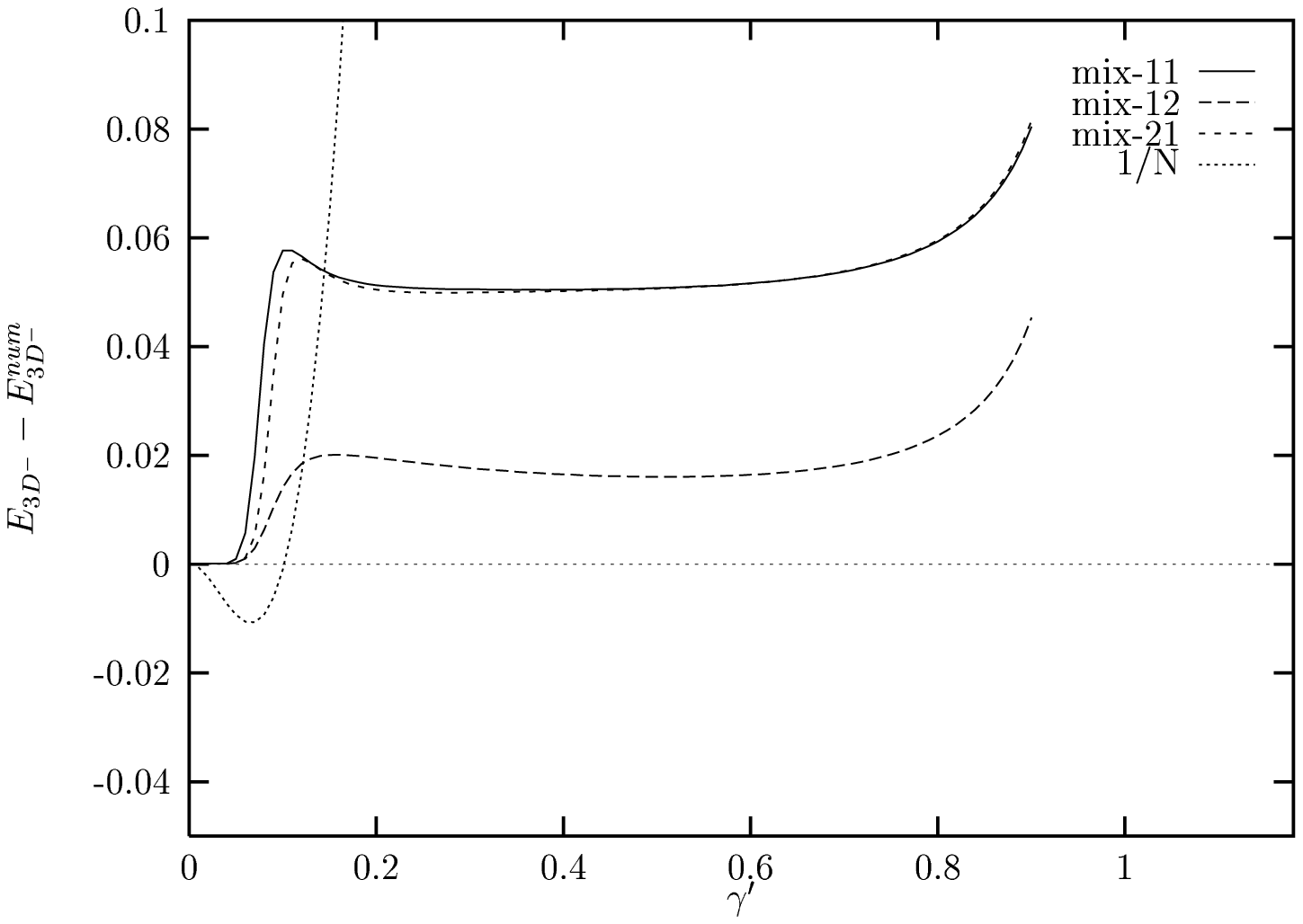,height=10cm}
\caption{The figure shows the difference between the numeric result for the 
$3D^-$ energy spectrum and the energy values computed with the help of 
the mix11 variational basis (solid line), mix21 variational (light dashed line), 
mix12 variational (heavy dashed line), and the shifted $1/N$ method (dotted line)}
\end{figure}

\section{Comparison with the 2D nonrelativistic Hydrogen Atom}
In this section we proceed to compare the results obtained for the energy spectrum of the 2D 
relativistic Hydrogen atom with those computed in the nonrelativistic limit, when the 2D Schr\"odinger equation was considered. In order to establish a better comparison we use numerical results obtained with the Schwartz interpolation method\cite{Schwartz}. In Tables 1 and 2 we exhibit 
different values of the energy for different magnetic field strengths. 

\begin{center}                  
{\scriptsize
\begin{tabular}{|l|l|l|l|} \hline                       
\multicolumn{1}{|c|}{$\gamma^{\prime}$} &\multicolumn{1}{|c|}{nonrel} &\multicolumn{1}{|c|}{rel} 
&\multicolumn{1}{|c|}{diff x 100} \\ \hline
0       &-0.4444444     &-0.44449574    &0.005134 \\ \hline
0.1     &-0.5239504     &-0.52421206    &0.026166 \\ \hline
0.2     &-0.5629132     &-0.5635173     &0.06041  \\ \hline
0.3     &-0.5635941     &-0.56468762    &0.109352  \\ \hline
0.4     &-0.5193465     &-0.52115542    &0.180892  \\ \hline
0.5     &-0.4095808     &-0.41248102    &0.290022  \\ \hline
0.6     &-0.1862031     &-0.19088784    &0.468474  \\ \hline
0.7     &0.2719562      &0.2640041      &0.79521   \\ \hline
0.8     &1.3504374      &1.33516298     &1.527442  \\ \hline
0.9     &5.1012487      &5.0599714      &4.12773   \\ \hline
\end{tabular}}

\par
\medskip
\end{center}
{\small{{\bf Table 1}  Relativistic energy values for  
$m=-1$, and a comparison with the non relativistic energy spectrum. The first column corresponds to the nonrelativistic energy, 
the second column is the relativistic energy $E-c^2$,  and the third column corresponds
to 100 times the difference between first and second columns
 }}                      

\begin{center}                  
{\scriptsize
\begin{tabular}{|l|l|l|l|} \hline                       
\multicolumn{1}{|c|}{$\gamma^{\prime}$} &\multicolumn{1}{|c|}{nonrel} &\multicolumn{1}{|c|}{rel} 
&\multicolumn{1}{|c|}{diff x 100} \\ \hline
0       &-0.16          &-0.1600366     &0.00366  \\ \hline
0.1     &-0.2605089     &-0.2606836     &0.01747  \\ \hline     
0.2     &-0.2731927     &-0.27364026    &0.044756  \\ \hline
0.3     &-0.2377636     &-0.2386137     &0.08501   \\ \hline
0.4     &-0.150434      &-0.1518818     &0.14478   \\ \hline
0.5     &0.0113883      &0.00901904     &0.236926  \\ \hline
0.6     &0.3009099      &0.29702154     &0.388836  \\ \hline
0.7     &0.8503051      &0.84361474     &0.669036  \\ \hline
0.8     &2.0734797      &2.06046406     &1.301564  \\ \hline
0.9     &6.1341553      &6.09847572     &3.567958  \\ \hline
\end{tabular}}
\par
\medskip
\end{center}
{\small{
{\bf Table 2} Relativistic energy values for  
$m=-2$, and a comparison with the non relativistic energy spectrum. The fist column corresponds to the nonrelativistic energy, the second column is the relativistic $E-c^2$ energy, 
and the third column corresponds
to 100 times the difference between first and second columns}
}                                
\par
\medskip

Tables 1 and 2 show that the role played by the relativistic corrections is to shift down the 
energy levels. The relativistic effects becomes noticeable when the magnetic field parameter $\gamma^{\prime}$  is close to unity. Figure 5. shows the dependence of this difference.   

\begin{figure}
\epsfig{file=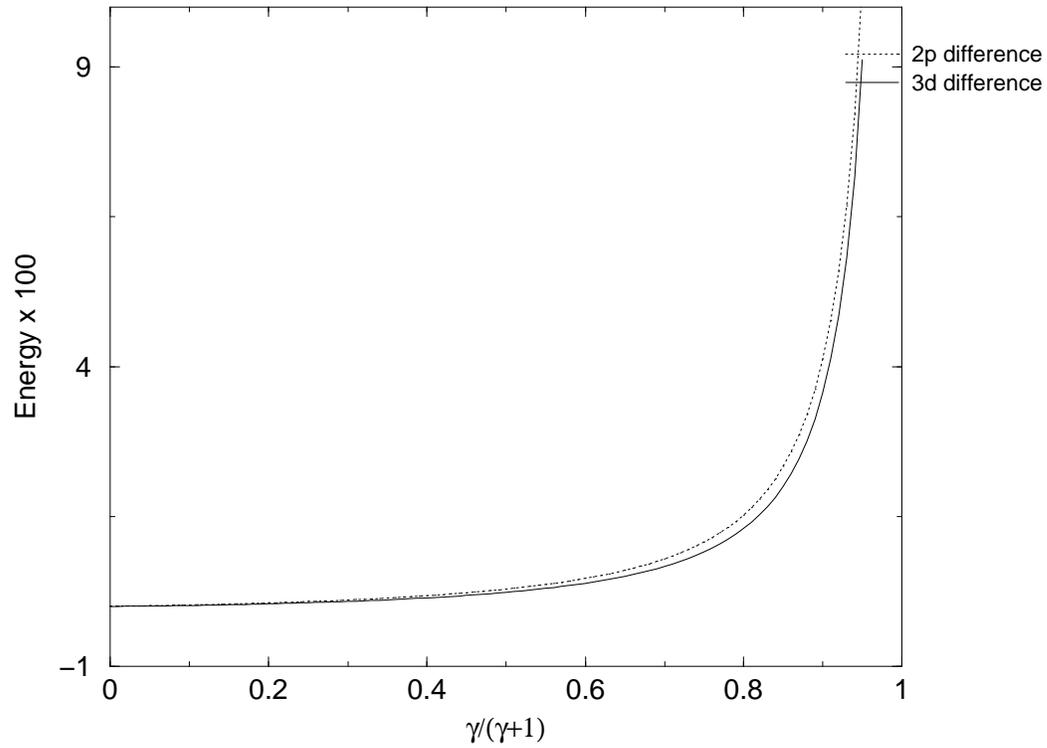,height=10cm}
\caption{The dotted line represents the difference between  nonrelativistic and relativistic $2p$ energy levels. The solid line represents the difference between nonrelativistic and relativistic $3d$ energy levels}. 
\end{figure}

\section{Concluding remarks}
Figures 1 thru 4 show that the variational mixed-basis method gives very
good results for the energy spectrum of a 2D relativistic Hydrogen atom in a 
constant magnetic field. The energy eigenfunctions obtained with this approach
give good energy values even for intermediate values of $\gamma^{\prime}$ as shown in Figs
1-4 where we compare the variational results with those obtained numerically 
via the Schwartz interpolation method \cite{Schwartz}. The shifted $1/N$ method
fails to give reasonably results when we apply its extension 
to the Klein Gordon equation \cite{Sever}. The advantage of the mixed variational method is that we obtain a simple form of the wave function and a reasonable good approximation without considering a large term variational basis. The role played by relativity consists in shifting down the energy levels as indicated in Figure 5. The results presented in this article were obtained considering a 
two-dimensional hydrogenic atom in the presence of a magnetic field 
perpendicular to the plane of motion. In this direction there are some 
differences between our approach and the method applied in Ref.
\cite{Chen1,Chen2}. 
Here the authors consider a non relativistic quasi two-dimensional system 
confined by a square-well $V_{B}(z)$. The Hamiltonian in Ref. \cite{Chen1,Chen2} contains 
this term and therefore the energy spectrum depends on the dimensions of the 
confining well. Since the variational technique applied in this paper is 
equivalent to the method suggested by Chen {\it et al} \cite{Chen1,Chen2}, 
 our results  
reduce, in the non relativisic limit, and when the width of the well is 
negligible,
 to those reported by Chen {\it et al}.
Finally we mention that  
$s$ states are not present for the 2D Klein-Gordon Hydrogen atom. 
The absence of $s$ $(m=0)$ states for the relativistic Klein-Gordon 2D Hydrogen atom can be understood if we look at the behavior of Eq. (\ref{7}) as $r$ approaches to zero. For m=0, we have a ``falling to center" problem\cite{Landau}, and this behavior is unobserved when we solve the 2D Dirac equation. A detailed discussion of relativistic effects including spin corrections will be presented in a future publication.  

\section*{Acknowledgments} We thank Dr. Ernesto Medina for reading and improving the manuscript. This work was supported by CONICIT under project 96000061.

\end{document}